\documentclass[12pt,a4paper]{amsart}
\newtheorem{thm}{Theorem}

\newtheorem{lem}{Lemma}
\newtheorem{df}{Definition}
\newtheorem{cor}{Corollary}

 \newcommand{\Real}{\mathbb R}
 \newcommand{\Nat}{\mathbb N}
 
 \newcommand{\Z}{\mathbb Z}

 \newcommand{\bs}{\bigskip}
 
 \newcommand{\eb}{\hfill q.e.d.}
 \newcommand{\tl}{{\mathcal T}}

 \newcommand{\A}{{\mathcal A}}

 \newcommand{\mix}{pointed pattern class}    
 \newcommand{\mixx}{doubly pointed pattern class}

 \newcommand{\Om}{\Omega}

 \newcommand{\Gr}{{\mathcal R}}
 \newcommand{\Gru}{\Gamma}

 \newcommand{\sst}{substitution}
 \newcommand{\CA}{$C^*$-algebra}

\newcommand{\mTxxx}{{\mathcal M}_{\rm I\!I\!I}}
\newcommand{\mTxx}{{\mathcal M}_{\rm I\!I}}
\newcommand{\mTx}{{\mathcal M}_{\rm I}}

\newcommand{\C}{{\mathcal C}}
\newcommand{\U}{{\mathcal U}}
\newcommand{\ein}{\succeq}
\newcommand{\nie}{\preceq}
\newcommand{\snie}{\succeq}
\newcommand{\sein}{\preceq}
\newcommand{\cp}{\vdash}
\newcommand{\AG}{almost-groupoid}
\newcommand{\rad}{\mbox{\rm rad}}
\newcommand{\erz}[1]{\langle{#1}\rangle}

\newcommand{\nct}{non commutative topology}
\newcommand{\N}{{\mathcal N}}
\newcommand{\svarphi}{\hat{\varphi}}

\newcommand{\spsi}{\hat{\psi}}

\newcommand{\id}{\mbox{\rm id}}
\newcommand{\bew}{{\em Proof:}}

\newcommand{\M}{\Gru}
\newcommand{\Mo}{\Gru_0}
\newcommand{\fol}[1]{(#1_n)_n}
\newcommand{\fl}[1]{\widetilde{#1}}
\newcommand{\MI}{\fl{\Gru}}
\newcommand{\AP}{approximating prehomomorphism}
\newcommand{\Li}{r}
\newcommand{\R}{d}

\newcommand{\isg}{\mathcal{ISG}}
\newcommand{\asg}{\mathcal{ASG}}
\newcommand{\Grt}{\Gr(\mTxx)}
\newcommand{\ldr}{local derivation rule}
\newcommand{\lm}{local morphism}
\newcommand{\LI}{{\mathcal L}}
\newcommand{\Cp}[1]{\C^c_{{#1}}}
\newcommand{\Gro}{\Omega}
\newcommand{\ac}{\cp}

\title[Topological equivalence of tilings]{Topological equivalence of tilings}

\author[J.\ Kellendonk]{Johannes Kellendonk\\
Fachbereich Mathematik, Technische Universit\"at Berlin}
\date{September 25, 1996}
\thanks{This work was supported by the DFG}
\address{Fachbereich Mathematik, Technische Universit\"at Berlin,
Stra\ss e des 17.\ Juni 136, 10623 Berlin, Germany.
E-mail: kellen@math.tu-berlin.de}

\begin{document}

\begin{abstract}
We introduce a notion of equivalence on tilings which is formulated in
terms of their local structure. We compare it with the known concept of
 locally deriving one tiling from another and show that two tilings
of finite type are topologically equivalent whenever their associated
groupoids are isomorphic.
\end{abstract}

\bibliographystyle{alpha}
\maketitle
\section*{Introduction}
In physics, tilings are used to model solids, in particular non-periodic ones.
Studying the possible types of long range ordered structures 
(and their implication on physical quantities)
amounts therefore in parts to the study of
(a suitable class of) tilings. In fact, the investigation of certain
tilings as idealized models for quasicrystals began
more than a decade ago so that one can find by now a large amount of
articles many of them being collected in \cite{OsSt,JaMo,AxGr}. 

Some elements of a theory of long range ordered structures
are based on tilings but require additional
information, for instance when it comes to 
the calculation of diffraction patterns 
(Fourier transforms).
Others depend only on the topological nature of the tiling, as e.g.\ the
$K$-theoretical gap labelling. Results of these are consequently more
qualitative in nature. The present article clearly belongs to the second
area. In particular, the specific shape or volume of the tiles which make 
up the tiling will not be of importance for us. Furthermore, due to the
locality of the interactions in the solid, it is only the
local structure of the tiling that matters, i.e.\ the way the tiling looks on
finite patches. 
One motivation to write this article is to illustrate that this local
structure can be described by an \AG\ resp.\ an 
inverse semigroup. The groupoid associated to the tiling arises
together with its topology functorially from the \AG.
%It may be seen as a non-commutative
%generalization of the topology of a space. 
The  algebraic structure
is defined on the most elementary level and therefore underlies the
construction of all topological invariants (including the group of possible
gap labels) of the tiling. 

The main aim of this article is however to give an answer to the question
under which circumstances two tilings give rise to isomorphic groupoids.
For that we introduce the notion of topological equivalence of tilings.
This notion is closely related 
to mutually locally derivability of the tilings,
a concept well known from physical considerations.\bs

The article is organized as follows.
We start with an informal description of the local structure of a tiling
as an example of an \AG\ resp.\ inverse semigroup.
After that we put this into a general context
and describe a functor which assigns to every \AG\
a groupoid with discrete orbits. We apply this to tilings,
obtaining the groupoid associated it, then emphasizing the particularities
of this case. 
By that we mean the existence of a metric structure 
which is well known for tilings and
with respect to which the functor looks like taking a closed
subspace of the metric completion. 
We compare the groupoid, which we also call in distinction the discrete
groupoid associated to the tiling, to the continuous groupoid, which is
often considered in the literature.  

In the next section we investigate the known concept
of local derivability of tilings which leads us to introduce
the notion of topological equivalence. Theorem~\ref{17093}
%and its corollary constitute 
constitutes the main result of this article.
It shows that topological equivalence of tilings -- a purely "local"
notion -- is sufficient and necessary for them to have isomorphic groupoids.
Whereas the metric structure is not used to define topological
equivalence of tilings the proof of theorem relies on this structure.
Everything is restricted to tilings which are of finite type.
The finite type (or compactness) condition which they satisfy is
the hypotheses for many compactness arguments.

In the final section we give a selected overview on topological invariants for
tilings. We mention the invariants of the groupoid-\CA, the $K$-groups,
and groupoid cohomology. But we will neither discuss the construction
of groupoid-\CA s (see \cite{Ren})
nor of its $K$-groups (see \cite{Bla}) nor of groupoid cohomology
(see \cite{Ren,Kum}).
We will also not illustrate the $K$-theoretical gap labelling 
but refer the reader to \cite{Be2,BBG,Ke2}.

\section{The local structure of a tiling}

In this article the following notion of tiling will be used.
A tile (in $\Real^d$) is a connected bounded subset of $\Real^d$ which is
the closure of its interior and may be decorated.\footnote{
The decoration, which may consist e.g.\ of arrows or colours, 
may serve for the purpose to distinguish 
translation classes of tiles which have
the same shape.}
A $d$ dimensional tiling is an infinite set of tiles which cover $\Real^d$ 
 overlapping at most at their boundaries.
A finite subset of a tiling is also called a pattern. 
Although often formulated for a specific tiling, the relevant quantities
like the groupoid associated to it and the \AG\ of its local structure depend
only on the congruence class of the tiling.
A tile- resp.\ tiling- 
resp.\ pattern-class shall here be an equivalence class under translation of a
tile resp.\ tiling resp.\ pattern, i.e.\ two such objects belong to the
same class if there is an $x\in\Real^d$ such that translation by $x$
applied to the tile resp.\ the elements of one set yield the other tile resp.\
the elements of the other. Note that a pattern class does not consist
simply of tile classes.

The local structure of a tiling is a multiplicative structure
determined by its pattern classes.
On the set of patterns of a given tiling one can easily introduce an
associative binary operation (multiplication), the union. 
But such an operation is not well defined on pattern classes. 
%of patterns (congruence w.r.t.\ to translations). 
In order to achieve this we need to
keep track of the relative position between patterns.
This can be done with the help of an additional choice of a tile in the 
pattern, such a composed object is called a pointed pattern.
Calling two pointed
patterns composable if their choice of tile coincides (in the tiling),
one may define an associative binary operation from the set of 
composable pairs 
into the set of pointed patterns as follows: the
union of the patterns of the composable pair yields the new pattern
and their common choice of tile the new choice.
This multiplication being only partially defined, it appears at first
sight to be a draw back, but its advantage lies in the possibility to
extend it to a well defined
partial multiplication on translation classes:
Call two \mix es composable if they have representatives which are composable
in the above sense, multiply them in case 
as above and take their translation class.

But this is not all we want. We want to be able to build arbitrary large 
pattern classes from a finite set of small ones using a multiplication.
This can obviously not be achieved by the above. Instead, if we look at
pattern classes with a choice of an ordered pair of tiles in it, calling that a
\mixx, we can make larger pattern classes from smaller ones as follows:
Ignoring the first resp.\ second choice in the ordered pair of the
 first resp.\ second pattern class we obtain two (simply) \mix es
which may be multiplied as above provided they are composable. Of the
resulting \mix\ we forget the choice of tile and take instead the
ordered pair which is given by the so far ignored tiles, namely the
first of the ordered pair of the first and the second 
of the ordered pair of the second pattern class we started with.
As we will elaborate below, this is a useful
algebraic structure which we call \AG.
Equivalently one could work, after adding a zero element, with inverse 
semigroups %\cite{Pet,Law}
\cite{Pet}.
We still keep the name \AG, because it is almost a groupoid and applying
a functor to it yields topological groupoids.
This functor is most natural in tiling theory since it furnishes tilings
from patterns.

\subsection{Almost-groupoids / inverse semigroups}
Let $\Gru$ be a set. A partially defined associative multiplication
is given by a subset $\Gru^\cp\subset \Gru\times\Gru$ of composable pairs 
(we write $x\cp y$ for $(x,y)\in\Gru^{\cp}$) with a
map $m:\Gru^\cp\to\Gru$ 
(we write $xy=m(x,y)$)
which is associative in the sense that, first,
$x\cp y$ and $xy\cp z$ is equivalent to $x\cp yz$ and $y\cp z$, and 
second, if $x\cp y$ and $xy\cp z$ then $(xy)z = x(yz)$.
Hence we don't have to care about brackets.

Relations or equations like the above in a set with 
partially defined associative multiplication make sense only if the
multiplications are defined, i.e.\
if the to be multiplied pairs are composable. 
In order to avoid cumbersome notation we
shall agree from now on that a relation involving products is true
if all multiplications involved are defined and it is then true.

Given such a set $\Gru$ with partially defined multiplication, suppose
that for some $a\in\Gru$ the equations
$axa=a$ and $xax=x$ were true for some $x\in\Gru$. Then $x$ is called an
inverse of $a$. $\Gru$ has a unique inverse (map) if any $a$ has a unique 
inverse. The inverse map is then denoted
by $a\mapsto a^{-1}$.
\begin{df}
An \AG\ is a set $\Gru$ with partially defined associative multiplication
and unique inverse. 
\end{df} 
A set with fully defined associative multiplication 
and unique inverse is an inverse semigroup, i.e.\
an inverse semigroup is an \AG\ for which 
$\Gru^\cp=\Gru\times\Gru$.  
In particular, adding a zero element $0$ to an almost groupoid $\Gru$ 
and extending the multiplication by $xy=0$ if $x\not\cp y$, and $0x=x0=0$,
yields an inverse semigroup with zero (which we write as $\Mo$).
Conversely, if $\Mo$ is an inverse semigroup with zero then
$\Gru=\Gru_0\backslash\{0\}$ with $\Gru^\cp=\{(x,y)|xy\neq 0\}$ is an
\AG. So we may apply the known results of inverse semigroup theory.
In fact, any statement below on \AG s may be reformulated as a statement
on inverse semigroups with zero element and vice versa. However, we find
the formulation in terms of \AG s more natural.

The elements of $\Gru^0:=\{x x^{-1}|x\in\Gru\}$ are called units. There are
the image of the frequently occurring maps
$\Li,\R:\Gru\rightarrow\Gru^0$ 
%(instead of $r,d$) 
given by $\Li(x) = xx^{-1}$ and $\R(x)=\Li(x^{-1})$ 
% occur over and over. 
Let us mention that the uniqueness of the inverse implies, first, that
units are the same as idempotents, i.e.\ 
$\Gru^0:=\{x\in\Gru| x^2=x\}$, and that they commute, 
and second, that the inverse map is an involution, 
in particular $(xy)^{-1}=y^{-1}x^{-1}$. A proof of that can be found in
\cite{Pet} formulated in the framework of inverse semigroups.

This has implications on which kind of elements are composable.
E.g.\ if $x\cp y$ then $(xy)^{-1}=y^{-1}x^{-1}$ so that we must also have
$ y^{-1}\cp x^{-1}$. Furthermore, under the same condition $x\cp y$ we have
$xy = xy y^{-1} x^{-1} xy$ so that we must have composabilities like
$xy\cp y^{-1}$ etc.. 
Similarly,
$\R(x)=\Li(y)$ implies $x=xyy^{-1}$ so that we must have $x\cp y$. 
If $x\cp y$ is even equivalent to $\R(x)=\Li(y)$, then $\Gru$ satisfies  
cancellation, i.e.
%$x\cp y$ and $x\cp z$ and 
$xy=xz$ implies $y= z$.
This is simply because for $xy=xz$ to be true we must have
$x\cp y$ and $x\cp z$. But then $y=\Li(y)y=\R(x)y=\R(x)z=z$.

%\noindent
%{\em Remark.} 

Note that a groupoid -- for an explicit definition c.f.\ \cite{Ren} --
is the same as an \AG\ which satisfies cancellation. 

The well known order relation
on inverse semigroups \cite{Pet} will be of great use here. One way of 
formulating it here is:
 \begin{df}
The order of an \AG\ is defined by\footnote{
We use here a direction of the order which coincides with the convention 
used in semigroup theory. It is reversed to that in \cite{Ke5}.}
\begin{equation}\label{29052}
x\sein y\quad\mbox{whenever}\quad  
r(x)=xy^{-1}.
\end{equation}
\end{df}
Note that $x\ein y$ is equivalent to $x^{-1}\ein y^{-1}$, and, if
moreover $y\cp z$, then $x\cp z$ and $xz\ein yz$. In other words  
the order is compatible with multiplication. Note also that
a groupoid has trivial order.
\begin{lem}\label{20091}
The set of all minimal elements of an \AG\ is a (possibly empty) ideal 
which is a groupoid.
\end{lem}
\bew\
Let $\Gru$ be an \AG\ and $x\cp y$ for two of its elements.
Suppose that $x$ is 
minimal and consider the
relation $xy\snie z$, i.e.\ $z^{-1}z=z^{-1}xy$.
We want to show that $z=xy$ and hence it is minimal. 
Since order is compatible with multiplication we have
$x\snie z y^{-1}$ hence $x= z y^{-1}$ by minimality. 
Since for units
$u$ holds $zu\sein z$ we conclude
$xx^{-1}=zy^{-1}yz^{-1}\sein zz^{-1}$, and
$xx^{-1}xy\sein zz^{-1}xy=z$ showing that $xy=z$. Thus $xy$ is minimal.
In particular, all minimal elements form an \AG\ (which may be empty). 
We want to show that it satisfies
cancellation, i.e.\ that $x\cp y$ implies $\R(x)=\Li(y)$.
If $x\cp y$ then $\R(x)\cp \Li(y)$ and hence
$\R(x), \Li(y)\ein \R(x)\Li(y)$. Minimality of 
$x$ implying that of $\R(x)$ and $\Li(x)$ shows that
$\R(x)=\Li(y)$.\eb\bs

Let $u\in\Gru^0$ and $c\in\Gru$. If $c\sein u$ then
$c^{-1}=d(c)u$ and in particular $c\in\Gru^0$. On the other hand
$u\sein c$ does for $u\in\Gru^0$ not have to imply that $c\in\Gru^0$.
But this latter property is useful in the sequel so that we give it a name.
\begin{df} 
An \AG\ is unit hereditary if, for $u\in\Gru^0$ and $c\in\Gru$,
$u\sein c$ implies $c\in\Gru^0$.
\end{df}
Either of the statements $xy^{-1}\in\Gru^0$ or $yx^{-1}\in\Gru^0$
implies that $x$ and $y$ have a lower bound (common smaller element).
E.g.\ if $xy^{-1}\in\Gru^0$ then $xd(y)$ is such a lower bound.
For a unit hereditary \AG\ the converse holds as well, namely
if $x$ and $y$ have a lower bound $z$ then $r(z)$ is smaller
than both, $xy^{-1}$ and $yx^{-1}$. Moreover, in that case
$z=zd(z)\sein xd(y)$. Therefore, if $\Gru$ is unit hereditary and $x$ and
$y$ have a lower bound then 
\begin{equation}\label{03061}
\max\{z\in\Gru|z\sein x,y\} = xd(y)
\end{equation}
and $r(x)y=r(y)x=yd(x)=xd(y)$.\bs

\noindent
{\bf Example 1.} Let $X$ be a topological space and $\beta_0(X)$ a 
(not necessarily proper) subset of the topology of $X$
which has the property that
any open subset of $X$ is a union of sets of $\beta_0(X)$
(i.e.\ it is a base of the topology)
and that it is closed under intersection. 
Then $UV=U\cap V$ defines a multiplication on $\beta_0(X)$.
Since the only solution of the equations $U\cap V\cap U=U$ and 
$V\cap U\cap V=V$
is given by $U=V$ and $U\cap U=U$,
$\beta_0(X)$ is a commutative inverse semigroup which consists of units
(idempotents) only. The empty set is a zero element in it and consequently
$\beta(X):=\beta_0(X)\backslash\{\emptyset\}$ a 
commutative \AG\ consisting of units only.
Its order is the inclusion of sets. Note that there are in general no
minimal elements in $\beta(X)$.\bs
 
\noindent
{\bf Example 2.} Let $\tl$ be a tiling of $\Real^d$. 
We already have explained in
words that the set $\mTxx$ of \mixx es 
carries a partially defined multiplication.
Let us reformulate this in more technical terms.
We start with defining an order relation on $\mTxx$, namely
$c\ein c'$ if $c'$ can be obtained from $c$
by addition of tiles but keeping the ordered pair of chosen
tiles fixed. Let $\mTxxx$ be the set of all pattern classes together with an
ordered triple of chosen tiles and denote for $\eta\in\mTxxx$  by
$\eta_{\hat{i}}\in\mTxx$ 
the \mixx\ which is obtained by forgetting the $i$th choice
in the triple.
Call two \mixx es $c,c'$ composable whenever there is
an $\eta\in\mTxxx$ such that 
$c\ein \eta_{\hat{3}}$ and $c'\ein \eta_{\hat{1}}$.
Then define the product of two composable elements
$$
cc' =  \max\{\eta_{\hat{2}}|\eta\in\mTxxx,c\ein \eta_{\hat{3}},
c'\ein \eta_{\hat{1}}\} 
$$
the maximum being taken with respect to the above order.
This defines an associative multiplication.
It turns out to have a unique inverse map $c\mapsto c^{-1}$ which is 
given by
interchange of the elements of the ordered pair of chosen tiles. 
Thus $\mTxx$ forms an \AG\ which is in general
not commutative. The order of the \AG\ coincides with
the order used to define composability.
In particular, the \AG\ of a tiling is unit hereditary. 
Note that there are no minimal elements in $\mTxx$.

A well known equivalence relation among tilings is that of two tilings
being locally isomorphic \cite{SoSt}. 
Thus are called two tilings which have the property that 
every pattern class of either
tiling can also be found in the other.\footnote{
The notion is used here in a stronger sense than in \cite{SoSt}
in that pattern classes are considered as 
equivalence classes under translations but not under rotations.}
This can here simply be expressed by saying that the
tilings lead to the same \AG.

Let $\mTx$ be the set of \mix es which are pattern classes together
with one chosen tile. We may identify $\mTx$ with the subset of $\mTxx$
consisting of those elements which are invariant under the inverse map,
i.e.\ for which the chosen tiles in the ordered pair coincide. 
Another specific property which holds for \AG s defined by tilings is that 
elements which are equal to
their inverse have to be units, i.e.\ under the above
identification $\mTx=\mTxx^0$. 

We shall be interested in tilings
which satisfy the following  finite type (or compactness) condition.
We call a pattern (and its class) connected if the subset it covers is
connected.
\begin{itemize}
\item
%Let $\mTzw\subset\mTxx\backslash\mTx$ be 
The set of connected \mixx es which consist of two tiles is finite.
\end{itemize}
Since tiles are bounded sets which have positive Lebesgue measure
this condition implies that, for any $r$, the maximal number of
tiles a pattern fitting inside an $r$-ball can have is finite.
From that one concludes that the above condition is 
equivalent to the requirement that the number of 
pattern classes fitting inside an $r$-ball is finite.
In particular $\mTxx$ is countable.

\subsection{From \AG s to groupoids}
We now aim at a functorial construction to obtain a 
topological groupoid from an \AG. For that we consider sequences
$(x_n)_{n\in\Nat}$ of elements $x_n\in\Gru$ which are
decreasing in that for all $n$: $x_n\ein x_{n+1}$. 
The set of all decreasing sequences, which 
is denoted by $\Gru^{\Nat}_{\ein}$,
carries a pre-order
\begin{equation}\label{18061}
\fol{x}\sein\fol{y}\quad\mbox{\rm
whenever}\quad \forall n\exists m:x_m\sein y_n.
\end{equation}
To turn this pre-order into an order one considers the
equivalence relation on $\Gru^{\Nat}_{\ein}$
\begin{equation}\label{18069}
\fol{x}\sim\fol{y}\quad\mbox{\rm
whenever}\quad \fol{x}\sein\fol{y}\quad\mbox{\rm
and}\quad \fol{y}\sein\fol{x}.
\end{equation}
On the set of equivalence classes, the elements of which we denote by 
$[\fol{x}]$,
\begin{equation}\label{10091}
[\fol{x}]\sein[\fol{y}] \quad\mbox{\rm
whenever}\quad  \fol{x}\sein\fol{y}
\end{equation}
is an order relation.
\begin{df}
For a given \AG\ $\Gru$, $\fl{\Gru}$ is the set 
$\Gru^{\Nat}_{\ein}$ modulo relation (\ref{18069}) and
$\Gr(\Gru)$ the set of minimal elements of $\MI$ with respect to the
order (\ref{10091}).
\end{df}
We identify the elements of $\Gru$ with constant sequences in $\MI$.
We use also the notation $\fl{x}$ for the elements of $\MI$.
\begin{lem}\label{21091}
If $\Gru$ is a countable \AG\ then any $x\in\Gru$ has a smaller
minimal element in $\MI$, in particular $\Gr(\Gru)\neq\emptyset$. 
\end{lem}
\bew\
Given $x\in\Gru$ there is a bijection
$\gamma:\Nat\to\Gru$ such that $\gamma(1)=x$. Now define
$\hat{\gamma}(1)=\hat{\gamma}(1)$ and
$\hat{\gamma}(n)=\hat{\gamma}(n-1)d(\gamma(n))$ if 
$\hat{\gamma}(n-1)\cp d(\gamma(n))$ and else
$\hat{\gamma}(n)=\hat{\gamma}(n-1)$. Then $(\hat{\gamma}(n))_n\in
\Gru^{\Nat}_{\ein}$. Now suppose that $\fol{y}\sein (\hat{\gamma}(n))_n$.
Then in  particular $y_m$ and 
$\hat{\gamma}(n)$ have for all $n,m\in\Nat$
a common smaller element. But this implies that
$\hat{\gamma}(\gamma^{-1}(y_m))\sein y_m$ and hence 
$\fol{y}\snie (\hat{\gamma}(n))_n$. Thus $(\hat{\gamma}(n))_n$, which is
certainly smaller than the constant sequence $x$, is a minimal element.
\eb\bs

Examples show that countability is not a necessary condition. 
\begin{lem}
$\MI$ is an \AG\ under the operations induced by point-wise operations on
 $\Gru^{\Nat}_{\ein}$, and its order coincides with the
order (\ref{10091}).
\end{lem}
\bew\
$\Gru^{\Nat}_{\ein}$ is an \AG\ under point-wise operations, i.e.\
composability is given by 
$\fol{x}\cp\fol{y}$ if $\forall n:x_n\cp y_n$ and then 
$\fol{x}\fol{y}=(x_ny_n)_n$, $\fol{x}^{-1}=\fol{x^{-1}}$.
Since order is compatible with multiplication 
$\fol{x'}\sim\fol{x}$ and $\fol{y'}\sim\fol{y}$ and $\fol{x}\cp\fol{y}$
imply, first $\fol{x'}\cp\fol{y'}$, and second $(x_ny_n)_n\sim (x'_ny'_n)_n$.
Furthermore $x\ein y$ being equivalent to $x^{-1}\ein y^{-1}$ implies that
$\fol{x}\sim\fol{y}$ is equivalent to $\fol{x^{-1}}\sim\fol{y^{-1}}$. 
From this follows the uniqueness of inversion. 
Hence also $\MI$ is an \AG. Its units are classes of sequences of decreasing
units of $\Gru$.
It is straightforward to see that its order is given 
by (\ref{10091}). \eb

\subsubsection{Morphisms of \AG s}
It turns out that the natural morphisms to look at in the context of
tilings are not homomorphisms but certain prehomomorphisms.
An order ideal of an \AG\ $\Gru$ is a subset $\N$ for which
$c\sein c'\in\N$ implies $c\in\N$.   
Any subset $\N$ generates an order ideal, namely
$I(\N)=\{x\in\Gru|\exists y\in\N:x\sein y\}$. 
%We shall deal with order ideals which are at the same time sub \AG s
%denoting that by \IO. 
We call an element of $\Gru^{\Nat}_{\ein}$ approximating if its class is
minimal.

\begin{df}
A prehomomorphism $\varphi:\Gru\to\Gru'$ between two \AG s is a map
which preserves composability, commutes with the inversion map, 
and satisfies for all
$x\cp y$
\begin{equation}\label{07031}
\varphi(xy)\nie\varphi(x)\varphi(y).
\end{equation} 
%A prehomomorphism of inverse semigroups with zero is called
%pure if $\varphi^{-1}(0)=0$.
A prehomomorphism is called approximating if it maps approximating
sequences onto approximating ones.
An \AP\ $\varphi:D(\varphi)\subset\Gru\to\Gru'$ is called a 
partial \AP\ or 
local morphism between $\Gru$ and $\Gru'$ if its domain $D(\varphi)$,
which is a sub-\AG\ of $\Gru$, 
is an order ideal.
\end{df}
\begin{lem}
Prehomomorphisms preserve the order.
\end{lem}
\bew\
%First, $\varphi(x)\sein\varphi(r(x))\varphi(x)$ shows that
%$\varphi(r(x))\sein\varphi(r(x))r(\varphi(x))$
%and hence $\varphi(r(x))=r(\varphi(x))$. 
$x\sein y$ is equivalent to $x=r(x)y$ and hence implies
$\varphi(x)\sein r(\varphi(x))\varphi(y)\sein\varphi(y)$.\eb\bs

This lemma implies that prehomomorphisms are
composable, and since the domain of $\psi\circ\varphi$, 
which is $D(\psi\circ\varphi)=\{x\in D(\varphi)|\varphi(x)\in D(\psi)\}$, 
is an order ideal of $\Gru$ local morphisms are composable as well.
A prehomomorphism $\varphi:\Gru\to\Gru'$ of \AG s can be extended to 
a prehomomorphism $\varphi:\Gru_0\to\Gru'_0$ of inverse semigroups
by simply setting $\varphi(0)=0$.
The condition that $\varphi:\Gru\to\Gru'$ 
preserves composability implies then for the extension that it satisfies
$\varphi^{-1}(0)=0$. Conversely, any
prehomomorphism $\varphi:\Gru_0\to\Gru'_0$ of inverse semigroups with zero
which satisfies $\varphi^{-1}(0)=0$ restricts to a prehomomorphism
on the \AG s.
A homomorphisms between \AG s is a prehomomorphism for
which (\ref{07031}) is an equality.

By element wise application to sequences, a prehomomorphism maps
decreasing sequences onto decreasing sequences, and 
moreover preserves equivalence classes.
Hence it extends to a 
prehomomorphism $\tilde{\varphi}:\MI\to{\MI'}$ 
through
\begin{equation}\label{20061}
\fl{\varphi}[\fol{x}]:=[(\varphi(x_n))_n].
\end{equation}
If $\varphi$ is a local morphism then we
denote by $\Gr(\varphi):\Gr(D(\varphi))\to\Gr(\Gru)$ 
the restriction of $\fl{\varphi}$ to the minimal elements $\Gr(D(\varphi))$. 

\subsubsection{Topology}

A topological \AG\ is an \AG\ which carries a topology 
such that the product and the inversion map are continuous,
$\Gru^\cp$ carrying the relative topology. 
A (locally compact) groupoid is called $r$-discrete if 
$r^{-1}(x)$ is discrete for any $x$,
or equivalently, if its set of
units $\Gru^0$ is open \cite{Ren}.

If nothing else is said $\Gru$ shall carry the discrete topology.
The topology of $\MI$ shall then be defined as the one which is generated by
$\beta_0(\MI):=\{\fl{U}_{x}|x\in\Gru_0\}$,
\begin{equation}
\fl{U}_{x} = \{\fl{y}\in\MI|\fl{y}\sein x\},
\end{equation}
$\fl{U}_0=\emptyset$, and 
$\Gr(\Gru)$ shall carry the relative topology, i.e.\
the one generated by 
$\beta_0(\Gr(\Gru))=\{\U_x,x\in\Gru\}$,
$\U_{x}=\fl{U}_{x}\cap\Gr(\Gru)$. 
Using set multiplication\footnote{For arbitrary subsets of 
$\MI_0$ is $\fl{U}\fl{V}=\{\fl{x}\fl{y}|\fl{x}\in \fl{U},
\fl{y}\in \fl{V},\fl{x}\cp \fl{y}\}$.} as multiplication on 
$\beta_0(\MI)$ resp.\ $\beta_0(\Gr(\Gru))$ we get: 
%$\beta_0(\MI)$ becomes an inverse semigroup with inclusion as order.
\begin{lem}
The maps $x\mapsto \fl{U}_x$ resp.\ $x\mapsto \U_x$
furnish an isomorphisms between the 
inverse semigroups $\Gru_0$ and $\beta_0(\MI)$ resp.\ $\beta_0(\Gr(\Gru))$.
\end{lem}
\bew\
%We first show that $\beta_0(\MI)$ is an inverse semigroup.
Let $x,y\in\Gru$. $\fl{U}_{x}\fl{U}_{y}\subset\fl{U}_{xy}$ follows
directly from the compatibility between order and multiplication
and $\U_{x}\U_{y}\subset\U_{xy}$ is then a consequence of Lemma~\ref{20091}. 
As for the converse, let 
$\fl{z}\sein xy$. 
Then first $x^{-1}\fl{z}\sein x^{-1}xy\sein y$,
and second $\fl{z}=r(xy)\fl{z}\sein r(x)\fl{z}$ hence $\fl{z}=x(x^{-1}\fl{z})$.
This shows that $\fl{z}\in\fl{U}_{x}\fl{U}_{y}$.
If moreover $\fl{z}$ is minimal then the factorization
$\fl{z}=(r(\fl{z}x))(x^{-1}\fl{z})$ shows  $\fl{z}\in\U_{x}\U_{y}$
as both, $r(\fl{z}x)$ and $x^{-1}\fl{z}$ are minimal. Thus
\begin{equation}\label{28051}
\fl{U}_{x}\fl{U}_{y}=\fl{U}_{xy}\quad\mbox{and}\quad\U_x\U_{y}=\U_{xy}.
\end{equation}
The considered maps are by definition surjective.
But either of $\fl{U}_{x}=\fl{U}_{y}$ or $\U_{x}=\U_{y}$
implies that $x\sein y$ and $y\sein x$ so that
the maps are injective as well.\eb\bs

If $\Gru$ is unit hereditary 
$\beta_0(\MI)$ and $\beta_0(\Gr(\Gru))$ are closed under intersection.
In fact,
$\fl{U}_{x}\cap\fl{U}_{y}\neq\emptyset$ whenever $x$ and $y$ have a lower 
bound in $\Gru$ and therefore
$\fl{U}_{x}\cap\fl{U}_{y}=\fl{U}_{r(x)y}$ (which might be empty)
and hence also $\U_{x}\cap\U_{y}=\U_{r(x)y}$.

\begin{thm}\label{12092}
$\Gr(\Gru)$ is an r-discrete topological groupoid whose topology is 
$T_1$. If $\Gru$ is unit hereditary then $\Gr(\Gru)$ is even Hausdorff.
\end{thm}
{\em Proof:}
$\U_x^{-1}=\U_{x^{-1}}$ is open showing continuity of the inversion.
By (\ref{28051}), 
$$m^{-1}(\U_x)=\bigcup_{(x_1,x_2)\in\Gru^\cp: 
x\ein x_1x_2}(\U_{x_1}\times\U_{x_2})\cap\Gr(\Gru)^\cp$$ is open as well
and hence multiplication continuous. 

Moreover, since $\R([\fol{x}])\nie \R(x_1)$ we have
$\Gr(\Gru)^0=\bigcup_{u\in\Gru^0}\U_u$ which is open and hence the groupoid
$r$-discrete. 

To show that $\Gr(\Gru)$ is $T_1$, i.e.\
that for all $\fl{x},\fl{y}\in\Gr(\Gru)$ with  $\fl{x}\neq\fl{y}$
there is an open $U$ containing 
$\fl{x}$ but not $\fl{y}$, let $\fol{x}$ resp.\ $\fol{y}$
be a representative for $\fl{x}$ resp.\ $\fl{y}$
observe that $\fl{x}\neq \fl{y}$ implies both, $\fl{x}\not\sein \fl{y}$
and $\fl{x}\not\snie \fl{y}$, and hence
the existence of an $n_0$ such that for all $n\geq n_0$:
$x_n\not\ein\fl{y}$ and $y_n\not\ein\fl{x}$.
Therefore any $U=\U_{x_n}$, $n\geq n_0$, does the job.

Now suppose that $\Gru$ is unit hereditary. We claim that
for some $m$, $x=x_{n_0}$ and $y_m$ do not have a smaller common element.
This then proves the Hausdorff property since 
for that $m$ is $\U_x\cap\U_{y_m}=\emptyset$ and 
$\fl{x}\in\U_x$ and $\fl{y}\in\U_{y_m}$.
To prove the claim suppose its contrary, i.e.\ $x$ and $y_m$ to have a
common smaller element for all $m$. Then $\fl{y}d(x)\sein\fl{y}$ which by
minimality implies $\fl{y}\sein\fl{y}d(x)$. 
In particular $\exists l\exists m:y_m\sein y_ld(x)$, 
and since $y_lx^{-1}$ is a unit $y_ld(x)\sein x$.
This contradicts the above.
\eb

\begin{thm}\label{29051}
Let $\varphi:D(\varphi)\to\Gru'$ be an 
\lm\ of \AG s and $\Gr(\varphi)$ be
the restriction of $\tilde{\varphi}$ to $\Gr(D(\varphi))$.
Then $\Gr(\varphi):\Gr(D(\varphi))\to \Gr(\Gru')$ is a continuous homomorphism
 between topological groupoids.
\end{thm}
{\em Proof:}
$\Gr(\varphi)$ is a prehomomorphism by construction. 
But cancellation implies that on groupoids the order is trivial and hence 
prehomomorphisms are homomorphisms.
To show continuity of $\Gr(\varphi)$
let $x'\in\Gru'$. Then
$\Gr(\varphi)([\fol{x}])\in\U_{x'}$ is equivalent to
$\exists n:x'\ein\varphi(x_n)$. 
Hence $\Gr(\varphi)^{-1}(\U_{x'})
\subset\bigcup_{y\in\Gru:\varphi(y)\nie x'}\U_y$.
Since also 
$\Gr(\varphi)(\U_y)\subset \U_{\varphi(y)}$,
and $x\sein y$ implies $\U_x\subset\U_y$, the above inclusion is in fact
an equality.
This shows continuity. \eb\bs

%Consider the category of \AG s with \lm s.
%Then it is clear that
%$\Gr(\id)=\id$ and $\Gr(\varphi\circ\psi)=\Gr(\varphi)\circ\Gr(\psi)$.
In fact, it is easily checked that $\Gr$ is a covariant functor of the 
category of \AG s with \lm s
into the category of $r$-discrete groupoids with
partial continuous homomorphisms.
Since $\Gr(\Gru)$ has trivial order decreasing
sequences are constant sequences. Therefore we may identify $\Gr\circ\Gr$
with $\Gr$.  

In general an \AG\ is non commutative. But in Example~1 we
have seen that bases %(without $\emptyset$)
of the topology of topological spaces 
which are closed under intersection give rise to 
\AG s which consist only of units so they are in particular commutative. 
Theorem~\ref{12092} and (\ref{28051})
show that any \AG\ may be identified (after adding
a zero element) with a base of the topology of a $T_1$ space
but only in the case where the \AG\ consists only of units 
its multiplication coincides with intersection.
In that case $\Gr(\Gru)=\Gr(\Gru)^0$. Hence if $x\cp y$, 
which means for groupoids $d(x)=r(y)$, then $y=x$.
In other words the groupoid operations are trivial, i.e.\
$x$ is composable only with itself, $x^2=x$, and $x^{-1}=x$.
So a topological groupoid which consists only of units is an ordinary
topological space.
This indicates why one may call the field to which this 
study of tilings belongs the \nct\ of tilings.

\subsubsection{Inverse semigroups of groupoids}
It is instructive to compare the inverse semigroup $\Gru_0$ from which we
obtained the groupoid with other
inverse semigroups which are often considered in connection with
groupoids. For instance in Renault's book \cite{Ren}
such inverse semigroups (with zero) are considered which consist of 
$G$-sets. A $G$-set is a subset $s$ of the groupoid $G$ 
which has the property that the restrictions of $r$ and $d$ to $s$
are both injective.
Multiplication is then given by set multiplication (and inversion applies
element-wise). The order is inclusion of sets and the empty set is the
zero element. If no more restrictions on the $G$-sets are given this is
called the inverse semigroup of the groupoid, we denote it here by $\isg(G)$.
In the context of $r$-discrete groupoids it is also interesting to look at
those $G$-sets which are compact and open. They form also a 
sub-inverse semigroup,
the ample semigroup of $G$ denoted here by $\asg(G)$. 
Note that both, $\isg(G)$ and $\asg(G)$ are closed under intersection.
Since the assignment of an inverse semigroup to the groupoid is reverse
to the functor $\Gr$ the natural question is whether they are somehow inverse
(leaving aside the more subtle question of how to assign to a 
groupoid homomorphism
a morphism of $\isg(G)$ or $\asg(G)$).
The answer is in general negative but we can say the following.
%(Since $r\circ\varphi=\varphi\circ r$ for a homomorphism
%of groupoids $\varphi$ (and the same with $d$) injectivity guarantees that
%$\varphi$, as a map between sets, maps $G$-sets onto $G$-sets but it is
%also necessary.)

The relation between the inverse semigroup $\Gru_0$ to start with and 
the inverse semigroups of $\Gr(\Gru)$-sets of $\Gr(\Gru)$ is rather obvious:
$\U_c$, $c\in \Gru$, is a $\Gr(\Gru)$-set so that
by (\ref{28051}) we may identify $\Gru_0$ as a
sub-inverse semigroup of $\isg(\Gr(\Gru))$. 
Furthermore, if the $\U_c$ are compact then $\asg(\Gr(\Gru))$
is given by (finite) unions of elements of $\Gru_0$ under this identification. 
But they are not equal (and not all finite unions are allowed). 

A topological space is called first countable if any point has a
countable local base. If it is $T_1$ then for any point $x$ 
there exists a descending sequence $\fol{U}$ of neighbourhoods such that 
$\bigcap_n U_n=\{x\}$. Such a sequence may be constructed as follows:
Let $\U(x)$ be a local base at $x$ and $\gamma:\Nat\to\U(x)$ be a bijection.
Define $\hat{\gamma}(1)=\gamma(1)$ and 
$\hat{\gamma}(n)=\gamma(m)$ where $m$ is the smallest number such that
$\gamma(m)\subset\gamma(n)\cap\hat{\gamma}(n-1)$.
This is a descending sequence and $x\in\bigcap_n \hat{\gamma}(n)$.
Let $y\neq x$ and $V$
be an open set containing $x$ but not $y$. Then there is a $U\in\U(x)$
such that $U\subset V$ and hence $y\notin \hat{\gamma}(\gamma^{-1}(U))$. 
Hence  $y\notin \bigcap_n \hat{\gamma}(n)$. 
Thus the $\hat{\gamma}(n)$ form the desired sequence. 

\begin{thm} If $G$ is a first countable groupoid whose topology is $T_1$ 
and generated by $\asg(G)$ then $\Gr(\asg(G)\backslash\{\emptyset\})=G$. 
\end{thm}
\bew\
Let $\Gru=\asg(G)\backslash\{\emptyset\}$.
The map $p([\fol{U}]=\bigcap_n U_n$ is easily seen to be a well defined map
from $\fl{\Gru}$ into the power set of $G$. 
Since the elements of $\asg(G)$ are compact $p([\fol{U}])$ is not empty.
Now suppose that $[\fol{U}]$ were minimal and $x\neq y$ both
in $\bigcap_n U_n$. 
Then there is a $V\in\asg(G):y\notin V,
x\in V$. It follows that $[\fol{V\cap U}]$ is strictly smaller 
than $[\fol{U}]$ which yields a contradiction. We conclude that
$p$ maps $\Gr(\beta(X))$ onto singletons.
Hence $p$ defines a map $p':\Gr(\beta(X))\to X$.  

Since $G$ is first countable and $T_1$ any point $x$
lies in the image of $p'$, namely according to the above remark
we can find a descending sequence of neighborhoods $(\hat{\gamma}(n))_n$
with $\{x\}=\bigcap_n\hat{\gamma}(n)$.
Now choose for any $n$ an $U'_n\in\asg(G)$ with
$x\in U'_n\subset \hat{\gamma}(n)$,  
and set $U_n=\bigcap_{i\leq n}U'_i$. Then $\fl{U}$ is a pre-image of $x$.
To show that $p'$ 
is injective suppose that $p([\fol{U^1}])=p([\fol{U^2}])$
so that $\fol{V}$ defined by $V_n=U^1_n\cap U^2_n$ is in $\fl{\Gru}$. 
Then $[\fol{V}]\sein
[\fol{U^i}]$ and by minimality $[\fol{V}]=[\fol{U^i}]$.
It is clear that 
${p'}^{-1}(U)=\U_U$ for $U\in\Gru$. Thus $p'$ is a homeomorphism. 
So it remains to show that $p'$ preserves composability and
$p'([\fol{U}][\fol{V}])=p'([\fol{U}])p'([\fol{V}])$, in case
$[\fol{U}]\cp[\fol{V}]$. 
Let $[\fol{U}]\cp[\fol{V}]$. This 
is equivalent to  $[(d(U)_n)_n]=[(r(U)_n)_n]$, and hence for
$\{x\}=\bigcap_n U_n$ and $\{y\}=\bigcap_n V_n$ it implies $x\cp y$.
Moreover, in that case 
%$xy\in \{x'y'|x'\in U_n,y'\in V_n,x'\cp y'\}$ for all $n$ and hence 
$xy\in\bigcap_n U_n V_n=p([\fol{U}][\fol{V}])$, and
since $p([\fol{U}][\fol{V}])$ is a singleton the claim follows.\eb

A topological space which has a base consisting
of closed (and open) sets is called zero dimensional. Hence the groupoid
of the last theorem is zero dimensional.
A zero dimensional $T_1$ space is totally disconnected, 
i.e.\ the only connected set containing a point $x$ 
is the singleton (one point set) containing $x$.

\subsubsection{The universal groupoid of an inverse semigroup}
The question of how to assign a groupoid 
$G(S)$ to an inverse semigroup $S$ in such
a way that $S$ may be identified with a sub inverse semigroup
of $\asg(G(S))$ has been thoroughly addressed in \cite{Pat1,Pat2}.
In particular, a construction is presented which yields the universal
groupoid $G_u(S)$ of an inverse semigroup $S$.
We have not made use of Paterson's approach
%as it didn't come to our attention early enough and appears to be more 
%involved. But we ought to
but followed different lines and therefore include a brief comparison
for completion.
This is best done by first presenting $\Gr(\Gru)$
in the manner it has been presented in \cite{Ke5} for tiling \AG s.

There is a right action of $\Gru$ on the space of units $\Gro=\Gr(\Gru)^0$ 
by means of partial homeomorphisms. Let 
\begin{equation} \nonumber
\Gro^\ac := \{(\fl{u},c)\in\Gro\times\Gru|r(c)\ein \fl{u}\}
\end{equation}
with relative topology, $\Gro\times\Gru$ carrying the product topology. Let 
$\gamma:\Gro^\ac\rightarrow \Gro:\,
(\fl{u},c)\mapsto d(\fl{u}c)$. 
Then $\gamma(\cdot,c):\U_{r(c)}\to\U_{d(c)}$ is 
%an action (because multiplication in $\MI$ is associative) which is 
a partial homeomorphism. 
%with domain $\U_{r(c)}$ and range $\U_{d(c)}$. 
Now consider the equivalence relation on $\Gro^\ac$ 
\begin{equation}
(\fl{u},c)\sim (\fl{u},c')\quad\mbox{whenever}\quad \exists n:
u_n c = u_n c'.
\end{equation}
It is straightforward to see that this definition is independent of the
choice of the representative $\fol{u}$ of $\fl{u}$ and that the relation is 
transitive. We denote the equivalence class of $(\fl{u},c)$ by
$[\fl{u},c]$.
\begin{lem}\label{21093}
Let $\Gr'(\Gru)$ be quotient of $\Gro^\ac$ by the above equivalence relation 
with quotient topology and consider the groupoid structure
defined by 
$[\fl{u},c][\fl{u}',c']=[\fl{u},cc']$ provided $\fl{u}'=d(\fl{u} c)$ and 
$[\fl{u},c]^{-1}=[d(\fl{u}c),c^{-1}]$.
Then $\Gr'(\Gru)$ is a groupoid which is isomorphic to $\Gr(\Gru)$. 
\end{lem}
\bew\
Let $f:\Gro^\ac\to \Gr(\Gru)$, $f(\fl{u},c):=\fl{u}c$. 
$f$ is surjective, since $\fl{c}=\fl{r(c)}c_1$ for some representative
$\fol{c}$ of $\fl{c}$. If $f(\fl{u},c)=f(\fl{u}',c')$ then first,
$\fl{u}=\fl{u}'$, and second $\exists n:u_n\sein r(c),r(c')$ so that
$(\fl{u},c)$ and $(\fl{u},c')$ are equivalent in the above sense.
The topology of $\Gr'(\Gru)$ is generated by sets of the form
$\left[\U_{u}\times\{c\}\cap\Gro^\ac\right]$. Such a set is equal
to $\left[\U_{r(uc)}\times\{uc\}\right]=\{[\fl{u},uc]|r(uc)\ein\fl{u}\}$
in case $u\cp c$ and otherwise empty.
Since
$f^{-1}(\U_c)=\U_{r(c)}\times\{c\}$ for any $c\in\Gru$, $f$ 
induces a homeomorphism between $\Gr'(\Gru)$ and $\Gr(\Gru)$.
It is straightforward to check that this homeomorphism preserves 
multiplication and inversion.\eb

To compare this with the universal groupoid $G_u$ defined by $\Gru_0$
\cite{Pat1,Pat2} we assume that $\Gru$ is countable.
Paterson looks a the space $X$ of all nonzero semicharacters of 
$\Gru^0_0$, i.e.\ at nonzero (inverse semigroup) homomorphisms 
$\alpha:\Gru^0_0\to\{0,1\}$,
the latter being a group under multiplication. 
Semicharacters yield an inverse semigroup under point-wise multiplication, but
$X$ not containing the zero map, it is an \AG\  
under point-wise multiplication. We denote by $1$ the semicharacter which
is identically to $1$. 
\begin{lem}
The map $\MI^0\to X\backslash\{1\}:\fl{u}\mapsto \alpha_{\fl{u}}$ where 
$\alpha_{\fl{u}}(v)=1$ if and only if $v\ein \fl{u}$ is an isomorphism
of \AG s (both containing only units).
\end{lem}
Let $\fl{u}\cp\fl{u}'$ for two elements of $\MI^0$. 
Then, for $v\in\Gru^0_0$, $\fl{u}\fl{u}'\sein v$ is equivalent to 
$\fl{u}\sein v$ and $\fl{u}'\sein v$. 
Hence $\alpha_{\fl{u}\fl{u}'}=\alpha_{\fl{u}}\alpha_{\fl{u}'}$. The above 
map is therefore a homomorphism which is obviously injective. 
Let $\alpha\in X$, $\alpha\neq 1$, and 
${\Gru^0}_\alpha=\{u\in\Gru^0_0|\alpha(u)=1\}$ be the support of $\alpha$.
${\Gru^0}_\alpha$ is a sub-inverse semigroup of $\Gru^0_0$ which is lower
directed, i.e.\ any two of its elements have a lower bound in it.
From the countability condition and Lemma~\ref{21091} follows that
$\fl{{\Gru^0}_\alpha}$ has a unique minimal element, call it $\fl{u}_\alpha$.
Then $\alpha=\alpha_{\fl{u}_\alpha}$. \eb\bs

Identifying $X$ with $\MI^0\cup\{1\}$, where we consider $1$ as an
extra element of $\MI$ which satisfies
$\forall\fl{c}\in \MI:1\fl{c}=\fl{c}=\fl{c}1$ and $11=1$,
Paterson's topology can be described as the one which is 
generated by sets of the form
\begin{equation}\label{25091}
A_{u;u_1,\cdots,u_k}:=A_u\cap A_{u_1}^{c}\cap\cdots
\cap A_{u_k}^{c}
\end{equation}
with $u,u_i\in\Gru^0_0$, $u_i\sein u$, $A_u=\fl{U}_u\cup\{1\}$, and
$A_{u_i}^c$ here denoting the complement of $A_{u_i}$. 
In particular, the relative topology of this topology on $\MI^0$ is finer then
the one we consider.

The universal groupoid $G_u(\Gru_0)$ is now obtained from a right action
of $\Gru$ on $X$.
Define 
\begin{equation}
X^\ac := \{(\fl{u},c)\in\MI^0\times\Gru|r(c)\ein \fl{u}\}\cup\{1\}\times\Gru
\end{equation}
with relative topology, $X\times\Gru$ carrying the product topology. 
Let $\gamma:X^\ac\rightarrow X$ with $\gamma(x,c) = d(xc)$. Again,
$\gamma(\cdot,c):A_{r(c)}\to A_{d(c)}$ is a partial homeomorphism.
Consider the equivalence relation on $X^\ac$ 
\begin{equation}
(\fl{u},c)\sim (\fl{u},c')\quad\mbox{whenever}\quad \exists n:
u_n c = u_n c',
\end{equation}
for $\fl{u}\in\MI^0$,
whereas $(1,c)$ is only equivalent to itself.
Again, it is independent of the
choice of representative and 
transitive. The universal groupoid 
$G_u(\Gru_0)$ is given by the quotient of $X^\ac$ 
w.r.t.\ the above equivalence relation 
with quotient topology and groupoid structure
defined by $[x,c][x',c']=[x,cc']$ provided $x'=d(xc)$ and 
$[x,c]^{-1}=[d(xc),c^{-1}]$, square brackets again denoting equivalence classes.
We will have more to say about the relation between $\Gr(\Gru)$ and 
$G_u(\Gru_0)$ in the case where $\Gru$ is a tiling \AG.

\subsection{Application to tilings} 
Let us see what $\Gr$ yields applied to the \AG\ $\mTxx$ of a tiling $T$.
%To visualize the elements of $\Grt$
For that
we consider a notion of radius of a \mixx. 
%and use it to define a metric.
%This notion will allows us to view $\MI$ as a metric space.
Let $\rad:\mTxx\to\Real^+$ be defined by
the Euclidean distance between the two tiles of the ordered pair
and the boundary of a pattern class\footnote{Choosing a representative
for the pattern class it is the Euclidean distance between
the boundary of the subset it covers and the subset covered by
the two tiles of the ordered pair.}. 
In particular $\rad(c)=\min\{\rad(\Li(c)),\rad(\R(c))\}$
and $c\sein c'$ implies $\rad(c)\geq\rad(c')$.
Furthermore, let $M_r(c)$, $r > 0$, 
be the \mixx\ which is obtained from $c$ by eliminating
all tiles which have distance greater than or equal to $r$ from both pointed 
tiles and $M_0(c)$ be the \mixx\ which is given by the pointed tiles only. 
%FALSCH: In particular, $M_{\rad(c)}(c)=c$.
The finite type (compactness) condition takes then the form
\begin{itemize}
\item
The set $\{M_r(c)|c\in\mTxx\}$ is finite for any $r$.
\end{itemize}
Of particular interest are \mixx es called $r$-patches which are those
which satisfy $c=M_r(c)$ and $\rad(c)\geq r$.
Consider the metric on $\Gru$ defined by
\begin{equation}\label{12091}
d(c,c')=\inf(\{e^{-r}|M_r(c)=M_r(c')\}\cup\{e^{-1}\})).
\end{equation}
\begin{thm}\label{16091}
Let $\mTxx$ be the \AG\ of a tiling which satisfies the finite type condition.
Then there is a continuous bijection between 
$\fl{\mTxx}$ and the metric completion %$\overline{\mTxx}$ 
of $\mTxx$ with respect to the above metric.
Furthermore, the sets $U_c$, $c\in\mTxx$ are metric-compact.
\end{thm}
\bew\
Let $\fol{c}$ be a decreasing sequence of \mixx es. 
Since $c\ein c'$ implies $M_r(c)\ein M_r(c')$ the finite type condition implies
the existence of an $N$ such that for all $n\geq N:M_r(c_n)=M_r(c_N)$. 
It follows that
$d(c_n,c_m)\leq e^{-r}$ for $n,m\geq N$, i.e.\ $\fol{c}$ is a 
Cauchy sequence.
Moreover, if $\fol{c}$ and $\fol{c'}$ are two decreasing sequences which are
equivalent in the sense (\ref{18069}) a similar argument shows that
$d(c_n,c'_n)\to 0$, i.e.\ that they are equivalent as Cauchy sequences.
Now fix an increasing sequence $(r_k)_k$ of positive numbers which diverges.
If $\fol{c}$ is a Cauchy sequence then
$\forall k\exists N_k\forall n\geq N_k:M_{r_k}(c_{N_k})=M_{r_k}(c_n)$.
Defining $j_k=j(\fol{c})_k=M_{r_k}(c_{N_k})$ yields thus a decreasing 
sequence for which $d(j_k,c_{N_k})\to 0$, i.e.\ which is equivalent to 
$\fol{c}$ as Cauchy sequence. 
%This shows that $\overline{\mTxx}$ is a quotient of $\fl{\mTxx}$.
Moreover, if $\fol{c}$ and $\fol{c'}$ are Cauchy equivalent sequences
then $j(\fol{c})=j(\fol{c'})$. If $\fol{c}$ is decreasing, then not only
$j(\fol{c})\ein \fol{c}$, but 
since $\forall n\exists k:r_k>\rad(c_n)$
also  $j(\fol{c})\nie \fol{c}$. So if $j(\fol{c})$ and $j(\fol{c'})$
are not equivalent
in the sense (\ref{18069}) they cannot belong to the same Cauchy class.
Therefore is the map %$\iota:\fl{\mTxx}\to\overline{\mTxx}$ 
which sends $\fl{c}$ to its Cauchy class a well defined bijection
between $\fl{\mTxx}$ and the metric completion of $\mTxx$.

To compare the topologies
%proof that the sets $U_c$ are compact we have to describe the
%extension of the metric to $\fl{\mTxx}$.
extend $M_r$ to $\fl{\mTxx}$ 
through $M_r(\fl{c})=\lim_nM_r(c_n)$, $[\fol{c}]=\fl{c}$.
The limit exists and is independent of the chosen representative 
by the same argument as above which in fact shows that 
$\lim_nM_r(c_n)=M_r(c_N)$ for some $N$. 
It is then straightforward to check that
the extension of the metric to the completion of $\mTxx$ is given by formally 
the same expression for $d$ as in (\ref{12091}). 
Again using the finite type condition one sees that
the image of the (continuous) function $d(\fl{c},\cdot):\fl{\mTxx}\to\Real^+$
is discrete apart from a limit point at $0$. Therefore 
$\epsilon$-neighbourhoods are closed and hence complete in the metric topology.
%In particular they are complete
%and since they are pre-compact there also compact. 
$\epsilon$-neighbourhoods are sets of the form 
$$U_r(\fl{c})=\{\fl{c}'|M_r(\fl{c}')=M_r(\fl{c})\}$$
(the smaller $\epsilon$ the bigger $r$)
but since $r$ is finite $U_r(\fl{c})=U_r(c_n)$ for some $n$
and representative $\fol{c}$. 
If $0<r_1<r_2$ then $U_{r_1}(c)=\bigcup_{c'|M_{r_1}(c')=M_{r_1}(c)}
U_{r_2}(c')$ but by the finite type condition only finitely many
sets in the union of the r.h.s.\ are mutually disjoint.
Thus for any $0<\epsilon_2<\epsilon_1$ holds that the 
$\epsilon_1$-neighbourhood has a finite cover by $\epsilon_2$-neighbourhoods,
i.e.\ $\epsilon$-neighbourhoods are pre-compact and hence compact.

If $c$ is an $r$-patch then
$U_r(c)=U_c$. For arbitrary $c\in\mTxx$ one has
$U_c=\bigcup_{c'\sein c} U_r(c')$ where $r$ is some number bigger
than the diameter of $c$ (the diameter of the set covered by a
representative of the pattern class in $\Real^d$). 
In particular, the
metric topology is finer than the original topology on $\fl{\mTxx}$.
But moreover, only finitely many
sets in the union of the r.h.s.\ are mutually disjoint
so that the $U_c$ are metric-compact.\eb\bs

To proceed let us extend the radius function 
$\rad:\fl{\mTxx}\to\Real^+\cup\{\infty\}$
through $\rad(\fl{x})=\lim_n \rad(x_n)$
the r.h.s.\ being independent of the representative.

\begin{lem} \label{18091}
Let $\mTxx$ be the \AG\ of a tiling which satisfies the finite type condition.
%Let $T$ satisfy the finite type condition.
$\fl{c}$ is minimal if and only if $\rad(\fl{c})=\infty$. Stated differently,
a sequence $\fol{c}\in{\mTxx}^\Nat_\ein$ is approximating 
if and only if the sequence $(\rad(c_n))_n$ diverges.
\end{lem}
\bew\
Suppose that $\rad(\fl{c})=R'<\infty$ and let $R>R'$.
There is at least one but at most finitely many $R$-patches $d_1,\dots,d_k$
for which $d_i\sein M_R(\fl{c})$.
Now consider the sequence which is obtained from a representative $\fol{c}$
of $\fl{c}$ by replacing each $c_n$ by $k$ elements
$r(d_1)c_n,\dots,r(d_k)c_n$.
Since $U_{c_1}$ is metric-compact 
the sequence has a metric-convergent subsequence, say $\fol{c'}$,
which we may assume to be decreasing (if not apply the map $j$ defined
in the proof of Theorem~\ref{16091}).
But then $\fol{c'}\sein\fol{c}$ and since $\rad(\fol{c'})\geq R$,
$\fol{c'}$ cannot be equivalent to $\fol{c}$. Hence $\fl{c}$ is not minimal.

For the converse suppose that $\rad(c_n)$ diverges and $\fol{c'}\nie\fol{c}$.
Since for all $n$ there is an $m$ such that $\rad(c_m)$ is larger
than the diameter of $c'_n$ 
this implies $c'_n\ein c_m$ and thus  
$\fol{c'}\ein\fol{c}$. Note that we do not need the finite type condition
for this part.\eb

\begin{lem} \label{20092}
Let $\mTxx$ be the \AG\ of a tiling which satisfies the finite type condition.
%Let $T$ satisfy the finite type condition. 
Then the relative topologies on $\Gr(\mTxx)$ coincide and
$\Grt$ is metric-closed in $\fl{\mTxx}$.
\end{lem}
\bew\
The relative metric-topology on $\Gr(\mTxx)$ is generated by sets
$U_r(\fl{c})\cap\Grt$ where $\rad(\fl{c})=\infty$. Hence 
$M_r(\fl{c})=M_r(c')$ for some $r$-patch $c'$ and thus $U_r(\fl{c})=U_{c'}$.
This shows that $U_r(\fl{c})\cap\Grt$ is open with respect
to the original topology on $\Grt$, i.e.\ the latter is finer
than the relative metric-topology. By Theorem~\ref{16091} 
the topologies coincide.

Now suppose that $\fl{x}$ is not minimal, i.e.\ $\rad(\fl{x})=R'<\infty$.
Let $R>R'$ and $\fl{y}$ be an element of the $e^{-R}$-neighbourhood
of $\fl{x}$. Then $\rad(\fl{y})=R'$ as well, 
and hence $\fl{y}$ is not minimal, i.e.\ $\fl{\mTxx}\backslash\Grt$
is metric open.\eb

\begin{cor}
%Let $\mTxx$ be the \AG\ of a tiling which satisfies the finite type condition.
Under the requirements of Theorem~\ref{16091} 
is $\U_c$ compact. In particular is
$\Gr(\mTxx)^0$ a compact zero dimensional metric space
and $\beta_o(\Grt)$ a sub-inverse semigroup of $\asg(\Grt)$.
\end{cor}

The compactness of $\U_c$ follows from Theorem~\ref{16091} and 
Lemma~\ref{20092}. Writing $\Gr(\mTxx)^0=\bigcup_u\U_u$, the union being
taken over all $u\in\mTx$ which consists only of one tile shows that
the finite type condition implies compactness for $\Gr(\mTxx)^0$.

Roughly speaking, we have shown that the elements of $\Grt$ can be seen
as limits of \mixx es whose radii eventually become infinite. 
This can be formulated as follows:
To a given approximating sequence $\fol{c}$ construct a covering of $\Real^d$
by first choosing a representative $\hat{c}_1$ for $c_1$ in $\Real^d$.
Then there are unique representatives $\hat{c}_n$ for $c_n$ such that
$\hat{c}_n$ is obtained from $\hat{c}_1$ by addition of tiles
(but keeping the ordered pair fixed).
Since $\rad(c_n)$ diverges 
$\bigcup_n\hat{c}_n$ is a covering of $\Real^d$ 
(each $\hat{c}_n$ is a set of tiles) together with an ordered pair 
of tiles. We call this a doubly pointed tiling. The elements of $\Grt$
are the classes of doubly pointed tilings which are obtained in this way.
The set of units $\Omega=\Gr(\mTxx)^0=\Gr(\mTx)$ 
can than be identified with classes of
tilings together with one chosen tile. It is called the hull of the tiling.

The relative Paterson topology on $\Omega$, c.f.\ (\ref{25091}), coincides
with the topology on $\Omega$ considered above, since 
the sets $\U_u$, $u\in\mTxx^0$, which generate the latter are closed. 
Moreover, since
$$\fl{U}_r(c)=\fl{U}_{M_r(c)}\backslash
\bigcup_{\mbox{\tiny $r$-patches }c'\neq c,c'\sein c}\fl{U}_{M_r(c')}$$
the relative Paterson topology on $\MI^0$ is finer than the metric
topology and hence $\Omega$ is a Paterson-closed subset of $X$.
It follows that $\Gr'(\mTxx)$ is a reduction of the
universal groupoid $G_u(\mTxx\cup\{0\})$ with respect to the 
subset $\Omega$, which fits well into the general theory of \cite{Pat2}. 

For later use we proof:

\begin{lem}\label{16092}
Let $\mTxx$ be the \AG\ of a tiling which satisfies the finite type condition.
Then any $\fl{c}\in\MI$ has a smaller minimal element.
\end{lem}
\bew\
Suppose that $\fl{c}$ is not minimal and therefore $\rad(\fl{c})=R<\infty$.
Fix an increasing diverging sequence of real numbers 
$(r_k)_k$ which are greater than $R$.
As in the proof of Lemma~\ref{18091} we construct $\fl{c}'_k$ such that
$\fl{c}'_k\sein \fl{c}$ and $\rad(\fl{c}'_k)\geq r_k$. 
Hence $\fl{c}'_k\in U_r(\fl{c})$ and since the latter is metric-compact 
the sequence  $(\fl{c}'_k)_k$ has a
metric-convergent subsequence 
converging to a class $\fl{c}'$ which is smaller than $\fl{c}$ and minimal.\eb

\subsubsection{A continuous groupoid associated to the tiling}
There is another topological groupoid one can assign to a tiling, which we
want to mention for comparison. Here one starts with the local isomorphism
class $\LI_\tl$ of a tiling $\tl$. This is the space of all tilings which
are locally isomorphic to $\tl$.
$\LI_\tl$ may be obtained as the closure
of the orbit of $\tl$ under the action of the group $\Real^d$ of translations
with respect to a metric. In fact, viewed as a geometrical object a tiling
may be translated, $\tl-x$, $x\in\Real^d$ is the covering given by
the sets $t-x:=\{y-x|y\in t\}$ where $t$ runs through all tiles of $\tl$.
Then $\LI_\tl$ is the closure of $\{\tl-x|x\in\Real^d\}$ under the metric
$$ d(T,T')=\inf(\{\{\epsilon |\exists x,x'\in 
\Real^d: r(T-x,T'-x')\geq \frac{1}{\epsilon},
|x|,|x'|<\epsilon\}\cup\{\frac{1}{\sqrt{2}}\}) $$
where $r(T,T')$ is the largest $r$ such that
$T$ and $T'$ agree on the $r$-ball around $0$ \cite{AP}.
The other groupoid which may now be assigned to $\tl$ is the transformation
group $\C_\tl:=\LI_\tl\times\Real^d$. 
Two of its elements $(T,x)$, $(T',x')$ are composable whenever 
$T'=T-x$ and then $(T,x)(T',x')=(T,x+x')$. The topology is the
product topology.
For distinction we call it the continuous groupoid assigned to the
tiling as opposed to the discrete one. How is it 
related to $\Gr(\mTxx(\tl))$?

Fix for each tile-class a point in its interior, we call it a puncture.
The punctures of the tiles of a tiling may be identified with a countable
subset of $\Real^d$.
Let $\Om_\tl$ be the 
subset of $\LI_\tl$ which consists of tilings with
the property that the puncture of one of its tiles 
identifies with $0\in\Real^d$.
The reduction of $\C_\tl$ by $\Om_\tl$, which is the sub-groupoid
$\{(T,x)\in\C_\tl|T,T-x\in\Om\}$, is the groupoid which has been
associated to an aperiodic tiling in \cite{Ke2}
It is isomorphic to  $\Gr(\mTxx(\tl))$,
an isomorphism is given by the map which assigns to $(T,x)$ the 
doubly pointed tiling class which is given by the class of $T$ and the
pair of tiles given by, first, the one which covers $0$, and 
second, the one which covers $x$.

Moreover, it has been proven by Anderson and Putnam \cite{AP}
that the above reduction of $\C_\tl$ is an abstract transversal of $\C_\tl$ 
in the sense of Muhly et al.\ so that by the work of the latter authors
\cite{MRW} the groupoid-\CA s of $\C_\tl$ and  $\Gr(\mTxx(\tl))$ are
stably isomorphic. 
  
\section{Topological equivalence and mutual local derivability}

If we focus on the role tilings play in solid state physics
when describing spatial structures, then several properties of the tiling
are unimportant. First of all, only the congruence class of the tiling
matters, and second, due to the locality of the interactions 
locally isomorphic tilings are equally well suited to describe that structure.
This can now all be taken into account by working with the \AG\ of the tiling.
However, investigating further the way how tilings model e.g.\ the 
arrangement of atoms (or ions) in solids one may take the point of view that
this should only be understood in a topological way. In particular
details like the precise position and strength of the bondings are to be added,
i.e.\ are not to be derived from the tiling.
This led Baake et al.\ from the theoretical physics group in T\"ubingen
to introduce another
equivalence relation between tilings which is based on mutual local 
derivability \cite{BSJ}, see also \cite{BaSc} for an overview.

Let $B_r(x)$ denote the closed
$r$-ball around $x$ and $B_r=B_r(0)$. Furthermore $\tl\sqcap B_r(x)$
is the pattern consisting of all tiles of $\tl$ which intersect with 
$B_r(x)$. %The following definition originates from \cite{BSJ}.
\begin{df}
$\tl_2$ is locally derivable from $\tl_1$ if there is an $r\geq 0$ such
that for all $x,y\in\Real^d$
\begin{equation}\label{12061}
(\tl_1-x)\sqcap B_r = (\tl_1-y)\sqcap B_r\quad\mbox{implies}\quad
(\tl_2-x)\sqcap \{0\} = (\tl_2-y)\sqcap \{0\}.
\end{equation}
\end{df}
Restricting our interest to tilings which satisfy the finite type condition
the knowledge of the correspondence between  
$(\tl_1-x)\sqcap B_r$ and $(\tl_2-x)\sqcap \{0\}$ for finitely many $x$
is enough to construct all tiles of $\tl_2$ from $\tl_1$. This obviously
defines a map $\ell:\LI(\tl_1)\to\LI(\tl_2)$, which is continuous, has
dense image and is therefore surjective. $\ell$ can be extended
to a surjective homomorphism of groupoids, $\ell:\C_\tl\to\C_{\tl'}$:
$(T,x)\mapsto (\ell(T),x)$.
We may call the replacement of
$\tl_1\sqcap B_r(x)$ by $\tl_2\sqcap \{x\}$
 a local derivation rule. 
In particular the above definition is equivalent
to saying that for all $r'\geq 0$ there is an $r\geq 0$ such
that for all $x,y\in\Real^d$
\begin{equation}\label{12062}
(\tl_1-x)\sqcap B_r = (\tl_1-y)\sqcap B_r\quad\mbox{implies}\quad
(\tl_2-x)\sqcap B_{r'} = (\tl_2-y)\sqcap B_{r'}.
\end{equation}
$\tl_1$ and $\tl_2$ are called mutually locally derivable 
%(written $\tl_1\sim_{mld}\tl_2$) 
if $\tl_2$ is locally derivable from $\tl_1$ and vice versa. This is an
equivalence relation which can be extended by saying that
$\tl_1$ and $\tl_2$ belong to the same MLD-class if there is a $\tl'_2$
which  is locally isomorphic to
$\tl_2$ and mutually locally derivable from
$\tl_1$. That this extension is an equivalence relation 
(in fact on LI-classes) follows from the observation that
if $\tl_2$ is locally derivable from $\tl_1$
and $\tl'_1$ is locally isomorphic to $\tl_1$ then the local derivation
rule can be used to locally derive a tiling 
$\tl'_2$ from $\tl'_1$. Then $\tl'_2$
%is by construction locally derivable from $\tl'_1$, and it 
has to be locally isomorphic to $\tl_2$. 
Moreover, the local derivation of $\tl_1$ from $\tl_2'$ yields the inverse 
of $\ell:\C_\tl\to\C_{\tl'}$ so that the latter becomes an isomorphism.
\begin{cor}\label{17092}
If $\tl$ and $\tl'$ are in the same MLD-class then the groupoids
$\Gr(\mTxx(\tl))$ and $\Gr(\mTxx(\tl'))$ 
are reductions (in fact abstract transversals) of the same groupoid.
In particular they have stably isomorphic groupoid-\CA s.
\end{cor}
This follows directly from the fact that $\C_\tl$ and $\C_{\tl'}$
are isomorphic and the above mentioned theorem of \cite{AP}.

The above corollary indicates that the T\"ubingen
formulation of local derivability is a good starting point to answer
the question under which circumstances $\Gr(\mTxx(\tl))$ and $\Gr(\mTxx(\tl'))$
are isomorphic.
In order to cast it
in a form applicable to our framework, using \AG s and
the discrete groupoid,
we are naturally led to strengthen and at the 
same time to generalize the concept of local derivation.
A strengthening comes along with the idea of preservation of the average 
number of tiles per unit volume
whereas a generalization is necessary
as we want to work in a purely topological setting.

\subsection{Constructing local morphisms from local derivation rules}
Suppose that $\N$ is a sub-\AG\ of $\Gru$
which is the order ideal generated by a finitely generated \AG, i.e.\ 
$\N=I(\erz{\C})$ where $\C$ is a finite set and $\erz{\C}$ the
\AG\ generated by it. 
Suppose furthermore that we have a map $\svarphi:\C\to\M'$ 
which satisfies conditions
which arise if it were the restriction of a prehomomorphism from $\N$
into an \AG\ $\Gru'$. A question which is of prime importance for sequel 
%the topological theory of tilings 
is whether we can construct  
a local morphism $\varphi:\N\to\Gru'$ from that map.

We call $n$ elements $c_1,\dots,c_n$ collatable if they may be composed,
i.e.\ if $\forall 1\leq k < n:c_1\dots c_k \cp c_{k+1}$.
Let $\C^{-1}=\C$ be a finite subset of an \AG\ and $\svarphi:\C\to\M'$
be a map into another \AG\ which satisfies for all $c,c_i\in\C$:  
\begin{itemize}
\item[E1] $\svarphi(c^{-1})=\svarphi(c)^{-1}$,
\item[E2] if
$c_1,\dots ,c_n$ are collatable then $\svarphi(c_1),\dots ,\svarphi(c_n)$
are collatable,
%\item if $c_i\in\C$ and $c_1c_2\in\C$ then 
%$\svarphi(c_1c_2)\sein\svarphi(c_1)\svarphi(c_2)$,
\item[E3] if
$c_1\dots c_n$ is a unit then 
$\svarphi(c_1)\dots \svarphi(c_n)$ is a unit.
\end{itemize}
Consider for $c\in\erz{\C}$
\begin{equation}
\Phi(c)=\{\svarphi(c_1)\dots \svarphi(c_n)|c_1\dots c_n=c,c_i\in\C\}.
\end{equation}
Since $c_1\dots c_n=c'_1\dots c'_{n'}$ implies that
$\svarphi(c_1)\dots \svarphi(c_n)(\svarphi(c'_1)\dots 
\svarphi(c'_{n'}))^{-1}$
is a unit any two elements of $\Phi(c)$ have a common smaller element,
i.e.\ $\Phi(c)$ is a lower directed set. 
Provided $\Phi(c)$ is finite we define
\begin{equation} \label{07034}
\varphi(c):=\min\Phi(c).
%=\bigwedge_{x\in\Phi(c)} x.
\end{equation}
Then $\varphi$  
commutes with the inverse map, because of
$\Phi(c)^{-1}=\Phi(c^{-1})$, and it satisfies inequality (\ref{07031})
since $\Phi(c_1)\Phi(c_2)\subset \Phi(c_1c_2)$. 
Thus $\varphi:\erz{\C}\to\M'$ is a prehomomorphism. 
If $H_\C(c):=\{c'\in\erz{\C}|c'\snie c\}$ has a unique minimal element
then $\pi:I(\erz{\C})\to \erz{\C}$: $\pi(c)=\min H_\C(c)$ is a 
prehomomorphism as well, and we may extend $\varphi$ through 
$\varphi\circ \pi$.

\begin{df}\label{18062}
We call a pair $(\varphi,\C)$,
where $\C=\C^{-1}\subset\Gru$ is finite and
$\svarphi:\C\to\M'$ satisfies conditions E1-3, 
a \ldr\ if it leads for all $c\in\Gru$ to finite lower directed sets 
$\Phi(c)$ and $H_\C(c)$ and $\varphi:I(\erz{\C})\to\Gru'$,
\begin{equation}\label{17091}
\varphi(c):=\min\Phi(\min H_\C(c))
\end{equation}
is approximating. 
\end{df}
\begin{lem}\label{13031}
Let $\mTxx$ and  $\mTxx'$ be two tiling \AG s. Suppose that
there exist a finite $\C=\C^{-1}\subset\mTxx$ and   
a map $\svarphi:\C\to\mTxx'$ which  
satisfies E1-3. Then $\Phi(c)$ and $H_\C(c)$ are finite lower directed sets.
\end{lem}
\bew\
$H_\C(c)$ is finite since any \mixx\ has only finitely many tiles.
It is lower directed since $\mTxx$ is unit hereditary.
As for $\Phi(c)$ we subdivide this set first into subsets
$c'\Phi_{c'c''}(c)c''$ where
$\Phi_{c'c''}(c):=\{
\svarphi(u_1)\dots\svarphi(u_n)|n\in\Nat,c=c' u_1\dots u_n c''\}$,
$u_i\in\erz{\C}^0$ and 
$c'=c'_1\cdots c'_k,c'_i\in\C$ none of the
$c'_i\cdots c'_j$, $1\leq i\leq j\leq k$, being a unit, and
the same conditions for $c''$.
Since there are only finitely many different 
units which satisfy $u \ein {c'}^{-1}c{c''}^{-1}$, units commute, and 
$\varphi(u) \varphi(u)=\varphi(u)$, $\Phi_{c'c''}(c)$ is finite.
Moreover, there are only finitely many different possibilities to choose
$c',c''$ so that $\Phi(c)$ is finite.\eb\bs

There is no reason why $\varphi$ should be approximating.

To connect the T\"ubingen formulation of local derivability with
the above and justify double use of the
word \ldr\ we proof:
\begin{thm}
Let $\tl'$ be locally derivable from $\tl$. Then there exists a
\ldr\ in the sense of Definition~\ref{18062},
$\svarphi:\C\subset\mTxx(\tl)\to \mTxx(\tl')$, such that
$\Gr(I(\erz{\C}))=\Gr(\mTxx(\tl))$ and
the induced homomorphism maps the class of $\tl$ onto that of $\tl'$.
\end{thm}
\bew\
First introduce punctures for the tile classes of $\tl$ 
which are chosen such 
that none of the punctures of tiles of $\tl$ lies on the boundary of
tiles of $\tl'$. 
For given $r'$ fix $r$ according to (\ref{12062}) and
let for any tile $t$ of $\tl$, $\hat{\ell}(t)=\tl'\sqcap B_{r'}(t^{pct})$,
where $t^{pct}$ is the puncture of $t$.
We now define a \ldr\ on the set $\Cp{r}$ of all $r$-patches $c$ for which
$M_0(c)$ is connected.
Let $m$ be a doubly pointed
pattern in $\tl$ of the class $\fl{m}\in\Cp{r}$. Denote the $i$th tile of its
ordered pair by $t_i(m)$.
Then $\svarphi(\fl{m})$ shall be the
class of the pattern 
$\hat{\ell}(t_1(m))\cup\hat{\ell}(t_2(m))$ with the ordered pair
$(\tl'\sqcap B_0(t_1(m)^{pct}),\tl'\sqcap B_0(t_2(m)^{pct}))$.
That $\svarphi(\fl{m})$ does not depend on the chosen representative 
$m$ for $\fl{m}$
is precisely the definition of local derivability. 
Defined in that geometrical way,
it is easy to see that $\svarphi$ satisfies the conditions E1-3. 
If $r'$ is larger than twice the diameter of the largest tile in
$\tl_1$ then $\hat{\ell}(t_1(m))\cup\hat{\ell}(t_2(m))$ is connected and
$\varphi$ approximating. By construction it maps the class of 
$\tl$ onto that of $\tl'$.
\eb\bs

Although the \ldr\ $\svarphi$ yields a homomorphism $\Gr(\varphi)$ which
is very similar to a restriction of the map 
$\ell:\C_\tl\to\C_{\tl'}$
constructed from the \ldr\ in the T\"ubingen version
it is neither injective nor surjective, in general. 
The geometrical picture of $\ell:\C_\tl\to\C_{\tl'}$ allows one to conclude
that $\Gr(\varphi)$ 
is surjective whenever the punctures for the tiles of $\tl$ can be chosen 
in such a way that
any tile of $\tl'$ contains at least one puncture.
(First, doubly pointed tiling classes $\fl{c}\in\Gr(\mTxx(\tl'))$ 
for which $r(\fl{c})$ is in the same class then $\tl'$ lie in the image
of $\Gr(\varphi)$, and then, by continuity, all of $\Gr(\mTxx(\tl'))$.)
Similarly, a necessary (but not sufficient)
condition for $\Gr(\varphi)$ to be injective is
that any tile of $\tl'$ contains at most one puncture.
Hence the failure of $\Gr(\varphi)$
to be an isomorphism may have its cause in that
the average number of tiles per unit volume is not preserved.

The converse of the theorem is false. 
If $\tl'$ is obtained from $\tl$ by a change of length scale 
or an overall rotation there would (apart from symmetric cases) not be a 
local derivation rule in the T\"ubingen sense but a local
derivation in the sense of Definition~\ref{18062} is 
is given by applying the change of length scale 
resp.\ rotation to the \mixx es.

\subsection{Topological equivalence}
An answer to the question under which circumstances two tilings lead to 
isomorphic groupoids shall be given here
in purely "local" terms, i.e.\ in terms
of \AG s and local derivation rules.
For that let us start with a lemma.
Let us use the notation that for subsets of an ordered set $X\sein Y$
if $\forall y\in Y \exists x\in X: x\sein y$.
\begin{lem}
Let $\varphi$ be a local morphism from a countable unit hereditary \AG\
$\Gru$ into itself. 
Then $\Gr(\varphi)=\id$ if an only if
$D(\varphi)\sein \Gru$ and $\varphi(c)$ and $c$ have 
for all $c\in D(\varphi)$ 
%which have a smaller minimal element
a lower bound.
\end{lem} 
\bew\
Suppose first that $\Gr(\varphi)=\id$ which in particular means
$\Gr(D(\varphi))=\Gr(\Gru)$. Let $c\in\Gru$, by Lemma~\ref{21091}
there is a smaller minimal element $[\fol{c}]$. 
It has a representative $\fol{c}$, $c_n\in D(\varphi)$. 
But then there exists already
some $c_n\in D(\varphi)$ for which $c_n\sein c$.
Furthermore, $\varphi(c_n)$ and $c_n$ must have for any $n$ a lower bound 
since they constitute equivalent sequences. Any such bound is also
a lower bound for $\varphi(c)$ and $c$.

As for the converse observe that under the assumption that 
$\varphi(c)$ and $c$ have a lower bound for all $c\in D(\varphi)$ we have
$\fl{\varphi}(\fl{c})d(\fl{c})\sein \fl{\varphi}(\fl{c}),\fl{c}$
and hence for minimal $\fl{c}$: $\fl{\varphi}(\fl{c})=\fl{c}$. Hence
$\Gr(\varphi)=\id|_{\Gr(D(\varphi))}$. 
But since $D(\varphi)$ is an order ideal,
$D(\varphi)\sein \Gru$ implies $\Gr(D(\varphi))=\Gr(\Gru)$.\eb

\begin{df}\label{17094} Two countable unit hereditary \AG s
$\Gru$ and $\Gru'$ are called topologically equivalent 
if there are \ldr s $\svarphi:\C\subset\Gru \to\Gru'$,
$\spsi:\C'\subset\Gru'\to\Gru$ such that for the induced
local morphisms $\varphi$ resp.\ $\psi$ holds 
$D(\psi\circ\varphi)\sein \Gru$, $D(\varphi\circ\psi)\sein \Gru'$,
and $\psi(\varphi(c))$ and $c$ have 
for all $c\in D(\psi\circ\varphi)$ resp.\  
$\varphi(\psi(c'))$ and $c'$ 
for all $c'\in D(\varphi\circ\psi)$ a lower bound.
Two tilings of finite type are called
topologically equivalent if their corresponding \AG s are 
topologically equivalent.
\end{df}
According to the above lemma the definition of topological equivalence may
equally well be formulated by saying that the local morphisms $\varphi$
and $\psi$ satisfy  $\Gr(\psi\circ\varphi)=\id$ on $\Gr(\M)$ and 
$\Gr(\varphi\circ\psi)=\id$ on $\Gr(\M')$.
By the functorial properties of $\Gr$ it implies that $\Gr(\Gru)$
and $\Gr(\Gru')$ are isomorphic and
shows at once that topological equivalence is indeed 
an equivalence relation.
According to Remark~1, being in the same MLD-class
is not sufficient to guarantee that the tilings are isomorphic.
It is sufficient only in case 
any tile of $\tl'$ contains exactly one of the punctures of $\tl$.
%(The map $\Gr(\psi\circ\varphi)$ obtained from $\ell$
%does not have to be the identity but may be an invertible translation.)
%In this sense topological equivalence is  stronger than mutual local
%derivability. But according to Remark~2 it is also more general.
  
\begin{thm}\label{17093}
Two \AG s of finite type tilings %which satisfy the finite type condition
are topologically equivalent whenever their associated
groupoids are isomorphic. 
\end{thm}
\bew\
We already mentioned above that topological equivalence implies
the existence of an isomorphism between the associated
groupoids.
For the converse let $f:\Gr(\mTxx)\to\Gr(\mTxx')$ be an isomorphism,
$\Gr(\mTxx)=\Gr(\mTxx(\tl))$, $\Gr(\mTxx')=\Gr(\mTxx(\tl'))$.
Let $Y\subset \Gr(\mTxx)$ resp.\ $Y'\subset \Gr(\mTxx')$
be the set of elements $y$ such that $M_0(y)$ is connected.
Furthermore, let $X=Y\cup f^{-1}(Y')$ and
$\C(r)=\{M_r(\alpha)|\alpha\in X\}$.
Since 
$f:\Gr(\mTxx)\to \Gr(\mTxx')$ is continuous and $X$ compact, 
\begin{equation} \label{07032}
\forall r'>0\exists r>0\forall \alpha\in X:
f(\U_{M_r(\alpha)})\subset\U_{M_{r'}(f(\alpha))}.
\end{equation} 
Choose $r>0$ and $r'>0$ satisfying (\ref{07032}), and define 
$\svarphi:\C(r)\to\mTxx'$ by
\begin{equation}
\svarphi(M_r(\alpha)) := M_{r'}(f(\alpha)).
\end{equation}
In particular  (\ref{07032}) implies 
\begin{equation} \label{07033}
f(\U_{c})\subset\U_{\svarphi(c)}
\end{equation} 
for all $c\in \C(r)$. To show that $\svarphi$ is a \ldr\ we first check
E1-3. E1 is clearly satisfied.
Using set multiplication and the convention that $\U_{cc'}=\U_0=\emptyset$
if $c\not\cp c'$ we obtain for collatable $c_1,\dots, c_n$
\begin{equation}\label{13032}
f ( \U_{c_1\dots c_n} ) = f (\U_{c_1})\dots f(\U_{c_n})\subset
 \U_{\varphi(c_1)\dots \varphi(x_n)}
\end{equation}
where we used (\ref{28051}) and that $f$ is a homomorphism of groupoids.
Therefore $ \U_{\varphi(c_1)\dots \varphi(x_n)}$
cannot be empty and hence $\varphi$ satisfies E2. 
To show E3 let $c_1\dots c_n$ be a nonzero unit. Then $f(\U_{c_1\dots c_n})
\subset\Gr(\mTxx)^0$. Since, for tilings, either 
$\U_c\cap\Gr(\mTxx)^0=\emptyset$ or $\U_c\subset\Gr(\mTxx)^0$ 
(\ref{13032})  implies E3 for $\varphi$. 
Therefore $\varphi$ extends to a prehomomorphism. Clearly 
$D(\varphi)=I(\erz{\C})\sein\mTxx$.
Moreover, (\ref{13032}) implies that (\ref{07033}) holds even for all 
$c\in I(\erz{\C})$. 
Therefore, if $\fol{c}$ is an approximating sequence, then
$f([\fol{c}])\in\bigcap_n \U_{\varphi(c_n)}$ or, stated differently,
$f([\fol{c}])\sein\fl{\varphi}([\fol{c}])$. Hence if 
$\varphi$ is approximating then $\Gr(\varphi)=f$.

So far we have only used that $f$ is a homomorphism. To show that
$\varphi$ is approximating we need to use its invertibility. 
Having nothing specific said about the choice of $r,r'$
we choose them now in a way that there exist 
$0< r_2\leq r$ and $r_1'\geq r'$ such that apart from (\ref{07032}) also holds
$f^{-1}(\U_{M_{r_1'}(\beta)})\subset\U_{M_{r}(f^{-1}(\beta))}$ and
$f^{-1}(\U_{M_{r'}(\beta)})\subset\U_{M_{r_2}(f^{-1}(\beta))}$
for all $\beta\in f(X)$.
Since $f(X)$ is compact as well this is possible. We then define
$\C'(r):=\{M_{r}(\beta)|\beta\in f(X)\}$, and
$\spsi_1:\C'(r'_1)\to\M$, $\spsi_2:\C'(r')\to\M$ by
\begin{equation}
\spsi_1(M_{r_1'}(\beta)) := M_{r}(f^{-1}(\beta))\:,
\quad
\spsi_2(M_{r'}(\beta)) := M_{r_2}(f^{-1}(\beta))
\end{equation}
for all $\beta\in f(X)$. 
Alike $\varphi$, $\psi_i$, $i=1,2$, extend to a prehomomorphisms
and $\Gr(D(\psi_i))=\Gr(\mTxx')$.
Moreover, $\svarphi\circ\spsi_1(M_{r_1'}(\beta))=M_{r'}(\beta)$
and $\spsi_2\circ\svarphi(M_{r}(\alpha))=M_{r_2}(\alpha)$
imply that $\psi_2\circ\varphi(c)\snie c$ for all 
$c\in D(\psi_2\circ\varphi)$ and 
$\varphi\circ\psi_1(c')\snie c'$ for all 
$c'\in D(\varphi\circ\psi_1)$. In particular,
$\Gr(\varphi\circ\psi_1)=\id$ on $\Gr(\mTxx')$
and $\Gr(\psi_2\circ\varphi)=\id$ on $\Gr(\mTxx)$.
Therefore, if $\fol{c}$ is an approximating sequence then
$\fl{\psi}_2([\fol{c}])=[(\psi_2\circ\varphi\circ\psi_1(c_n))_n]
\ein [\psi_1(c_n))_n]$.
In particular, if next to $\fl{c}$ also $\fl{\psi}_2(\fl{c})$
is minimal then 
$\fl{\psi}_2(\fl{c})=\fl{\psi}_1(\fl{c})$.
Now let $\fl{c}\in\Gr(\mTxx)$. 
By Lemma~\ref{16092} there is a $\fl{c}'\in\Gr(\mTxx')$ with
$\fl{c}'\nie\fl{\varphi}(\fl{c})$. Then 
$\fl{\psi}_2(\fl{c}')\nie\fl{\psi}_2\circ\fl{\varphi}(\fl{c})=\fl{c}$, i.e.\
$\fl{\psi}_2(\fl{c}')$ is minimal, and hence $\fl{c}=\fl{\psi}_1(\fl{c}')$,
and consequently 
$\fl{\varphi}(\fl{c})=\fl{c}'$.
It follows that $\fl{\varphi}$ is approximating and hence $\Gr(\varphi)=f$.
But then the above implies that $\psi_i$, $i=1,2$, are approximating and
$\Gr(\psi_i)=f^{-1}$. Hence $\svarphi$ and $\spsi_i$ for either of the
$i=1,2$ satisfy according to Lemma~\ref{21091} the requirements of
the definition of locally topological equivalence.\eb\bs

In fact, we have proven a little more, namely that any isomorphism
between groupoids associated to finite type tilings is "locally defined",
i.e.\ it can be obtained by a local derivation rule.
One could also define a stronger form of topological equivalence between
two tilings $\tl$, $\tl'$ in that one requires in addition for the local
morphism of Definition~\ref{17094} that $\Gr(\varphi)$
maps the class of $\tl$ onto that of $\tl'$. 
This is then equivalent to the existence of an isomorphism between
$\Gr(\mTxx(\tl))$ and  $\Gr(\mTxx(\tl'))$ which 
maps the class of $\tl$ onto that of $\tl'$. 

A simple example for which the construction
of a prehomomorphism of the theorem can be carried out, not yielding an
approximating one, is
the constant map $f:\Gr(\mTxx)\to \Gr(\mTxx')$ given by
$f(\fl{c})=\fl{u}$, $\fl{u}\in\Gr(\mTxx')^0$ fixed. 
The above construction yields $\varphi(c)=M_{r'}(\fl{u})$ for all
$c\in D(\varphi)$ which is not approximating.\bs

\section{A selected overview on topological invariants of tilings}

We have shown that the topological groupoid $\Gr(\mTxx)$ is a
complete invariant for a topological equivalence class of tilings
which are of finite type.
This answers the question %(on the mathematical side)
under which circumstances two tilings of finite type lead to the same groupoid.
Furthermore it means that the groupoid 
contains all physically interesting topological
information about a tiling, the
prime example of that being the $K$-theoretic gap labelling.

%It is therefore natural to ask whether we can classify all groupoids
%coming from tilings. 

The question immediately following such a result is that after an
invariant for tiling-groupoids which is computable
and distinguishes between non-isomorphic groupoids (the term
invariant always referring to a quantity which depends on
isomorphism classes). In fact, the determination whether two
such groupoids are isomorphic or not can be rather difficult,
and what we have in mind here is something like Elliot's classification of 
$AF$-algebras by means of their scaled ordered $K_0$-group \cite{Ell1}.
These groups may be in many cases easily computed \cite{Eff}.
So one might hope that the $K$-theory of the groupoid-\CA\ is a good
starting point to classify all groupoids coming from tilings. 
And in fact, if one restricts its attention only to
the groupoid-\CA\ of the groupoid, then, for $1$-dimensional tilings -- which 
may be viewed as topological dynamical systems -- 
one obtains a \CA\ which is the limit of circle algebras. Elliot's
classification extends to such algebras \cite{Ell2},
the scaled ordered $K_0$-group
of the groupoid-\CA\ is a complete invariant as well.
A full treatment of the one dimensional case including an interpretation in
dynamical terms can be found in \cite{HPS,GPS}.
In higher dimensions, it is not yet clear whether $K$-theory yields complete
invariants for the groupoid-\CA s of tilings but the ordered $K_0$-group is
still an interesting object to consider, after all it has physical
signification in the gap-labelling.
However, it should be said that there are non-isomorphic tiling
groupoids which give rise to isomorphic \CA s, so that the $K$-theory
of the latter cannot be a complete invariant for tiling groupoids.
It is known that groupoids are invariants for pairs of \CA s,
the groupoid-\CA\ and a Cartan subalgebra of it \cite{Ren}.

\subsection{K-theoretic invariants}
The definition of the (reduced or full) groupoid \CA\
of an $r$-discrete groupoid can be found in \cite{Ren} or,
in the special context of tilings, in \cite{Ke2,Ke5}.
In the latter case, it may be seen as the $C^*$-closure of a representation of
the inverse semigroup $\mTxx\cup \{0\}$ by means of partial
symmetries of a Hilbert space and coincides with the algebra
of observables for particles moving in the tiling.
To be more precise, a priori on distinguishes two such closures, obtaining
the reduced or the full algebra. But since the (discrete) groupoid of
a tiling is the abstract transversal of a transformation group with 
amenable group, its reduced and full groupoid-\CA\ coincide \cite{MRW,Muh}.

The $K$-theoretic invariants of the groupoid-\CA\ 
$\A_\tl$ of $\Gr(\mTxx(\tl))$ are topological invariants
of the tiling. 
The results which could be obtained so far are, apart from
periodic tilings, all related to tilings which are invariant under
a primitive invertible \sst.
For one dimensional tilings the $K$-theory is computed 
in \cite{For,Hos}. For  higher  dimensional tilings
the (integer) group of coinvariants (which is actually
a cohomology group) together with
a natural order could be obtained in \cite{Ke5}. 
For tilings which allow for a locally defined $\Z^d$-action, $d\leq 3$, the
group of coinvariants embeds as ordered unital group into the $K_0$-group.
This is enough to solve the $K$-theoretical gap-labelling for these.
Explicit calculations include Penrose tilings \cite{Ke5} and
octagonal tilings \cite{Ke6}. Further results are obtained in terms of
cohomology groups, see below. 

But before coming to that let us recall Corollary~\ref{17092} which has
as a consequence that $K_1$-groups and ordered $K_0$-groups alone (without 
order unit) are invariants for MLD-classes of tilings.
That the order unit may distinguish elements of such a class may be seen
from the cases in which $\Gr(\varphi)$ is injective but not surjective.
In particular, any tile of $\tl'$ contains at most one puncture of a tile
of $\tl$ but some of them carry none. In this situation one can
identify $\A_\tl$ with a full corner of $\A_{\tl'}$ and the induced
order isomorphism between the ordered $K_0$-groups maps the order
unit of $K_0(\A_\tl)$ onto an element which is strictly smaller than
the  order unit of $K_0(\A_{\tl'})$ \cite{Ke5}.

\subsection{Cohomological invariants}
Another topological invariant of a groupoid is its cohomology. 
If one considers cohomology groups of the discrete groupoid
with integer coefficients then, at least 
for tilings which carry a local $\Z^d$-action, unordered $K$-groups are
isomorphic to cohomology groups \cite{FoHu}. E.g.\ the non-vanishing
cohomology group of highest degree, which is the group of coinvariants,
is a direct summand of the $K_0$-group. This was taken advantage of 
already above.

On the other hand Anderson and Putnam showed that unordered $K$-theory of
two dimensional \sst\ tilings is isomorphic to the Czech-cohomology
of a certain CW-complex \cite{AP}. 
They computed the latter for a number of tilings
including Penrose tilings. In particular they obtained as well
the $K_1$-group. The route they took is different from the one in 
\cite{Ke5}, but the actual calculations, as far as they concern
the common part of the results, reduce
at the end in both cases to the computation of images and kernels of 
combinatorial matrices, which are almost the same.

Comparing the types of invariants it can be said that 
cohomology groups give a finer grading than $K$-groups
but a priori no order. 
This is a severe draw back due to the vast possibilities
of orders on such groups. In particular, integer valued cohomology is
not a complete invariant for tilings either.\bs

Other cohomology groups of groupoids are also of interest for physics.
The second cohomology group of a groupoid with coefficients in the
circle group provides the twisting elements for the construction of
the twisted groupoid-\CA\ \cite{Ren}. 
For the simpler case of the group $\Z^2$ (which is of course a groupoid)
the twisted group-\CA\  is very important.
It is an irrational rotation algebra which is the observable e.g.\ for
for particles which move on the lattice $\Z^2$ (a periodic tiling) 
and which are subject to a constant perpendicular magnetic field 
\cite{Zak,TNK,Be1}. %,FrKo}.
The flux through the unit cell (a tile)
may be interpreted as the cocycle which yields the twisting element. 
It would therefore be rather interesting to compute the full second cohomology
group with coefficients in the circle group for non periodic tilings.

%\bibliography{Lit160996}

\end{document}